\documentclass[twocolumn]{jpsj2} 
\title{
Orbital Configurations and Magnetic Properties of Double-Layered Antiferromagnet Cs$_3$Cu$_2$Cl$_4$Br$_3$
}

\author{
Fumiko \textsc{Yamada}$^1$\thanks{E-mail: yamada@lee.phys.titech.ac.jp.}, Toshio \textsc{Ono}$^1$, Hidekazu \textsc{Tanaka}$^1$ and Jun-ichi \textsc{Yamaura}$^2$
}

\inst{
	$^{1}$Department of Physics, Tokyo Institute of Technology, Oh-okayama, Meguro-ku, Tokyo 152-8551\\
	$^{2}$Institute for Solid State Physics, The University of Tokyo, 
Kashiwanoha, Kashiwa, Chiba 277-8581
}

\abst{
We report the single-crystal X-ray analysis and magnetic properties of a new double-layered perovskite antiferromagnet, Cs$_3$Cu$_2$Cl$_4$Br$_3$. This structure is composed of Cu$_2$Cl$_4$Br$_3$ double layers with elongated CuCl$_4$Br$_2$ octahedra and is closely related to the Sr$_3$Ti$_2$O$_7$ structure.
An as-grown crystal has a singlet ground state with a large excitation gap of $\Delta/k_{\rm B}\simeq 2000$ K, due to the strong antiferromagnetic interaction between the two layers. Cs$_3$Cu$_2$Cl$_4$Br$_3$ undergoes a structural phase transition at $T_{\rm s}\simeq330$ K accompanied by changes in the orbital configurations of Cu$^{2+}$ ions. Once a Cs$_3$Cu$_2$Cl$_4$Br$_3$ crystal is heated above $T_{\rm s}$, its magnetic susceptibility obeys the Curie-Weiss law with decreasing temperature even below $T_{\rm s}$ and does not exhibit anomalies at $T_{\rm s}$. This implies that in the heated crystal, the orbital state of the high-temperature phase remains unchanged below $T_{\rm s}$, and thus, this orbital state is the metastable state. The structural phase transition at $T_{\rm s}$ is characterized as an order-disorder transition of Cu$^{2+}$ orbitals.  
}

\kword{Cs$_3$Cu$_2$Cl$_4$Br$_3$, double-layered perovskite, Sr$_3$Ti$_2$O$_7$ structure, Jahn-Teller effect, singlet ground state, structural phase transition, orbital order and disorder}

\begin{document}
\maketitle

\section{Introduction}
The nature of the exchange interaction between magnetic ions is strongly correlated to their orbital configurations \cite{Kanamori}. This is particularly significant when the magnetic ions are Jahn-Teller active such as Cu$^{2+}$, Cr$^{2+}$ and Mn$^{3+}$. For example, in KCuF$_3$ with a perovskite structure, orbital ordering leads to a one-dimensional antiferromagnetic chain \cite{Satija}. In this paper, the orbital configurations of Cu$^{2+}$ ions and the resulting magnetic ground states in Cs$_3$Cu$_2$Cl$_4$Br$_3$ are discussed. As shown later, this copper compound crystallizes in the double-layered perovskite structure, which is closely related to the Sr$_3$Ti$_2$O$_7$ structure \cite{Sr3Ti2O7}. 

Cs$_3$Cu$_2$Cl$_4$Br$_3$ was accidentally synthesized in the process of preparing a mixed crystal of CsCuCl$_3$ and CsCuBr$_3$. 
Cs$_3$Cu$_2$Cl$_4$Br$_3$ is a new compound that has not been reported to date. In this work, we carried out the X-ray crystal structural analysis and magnetic measurement of Cs$_3$Cu$_2$Cl$_4$Br$_3$. We found that Cs$_3$Cu$_2$Cl$_4$Br$_3$ has a double-layered structure of the Sr$_3$Ti$_2$O$_7$ type and that the magnetic ground state of an as-grown crystal is a spin singlet with a large excitation gap of $\Delta/k_{\rm B}\simeq 2000$ K, due to the strong dimerization between two layers.
The double-layered structure related to the Sr$_3$Ti$_2$O$_7$ structure has often been observed in cupric oxides, many of which exhibit high-$T_{\rm c}$ superconductivity \cite{YBCO,Bi2212,TlCaBaCuO}. In cupric halides, only K$_3$Cu$_2$F$_7$ is known to have the double-layered structure of the Sr$_3$Ti$_2$O$_7$ type \cite{K3Cu2F7}. Cs$_3$Cu$_2$Cl$_4$Br$_3$ shows another example of the Sr$_3$Ti$_2$O$_7$ structure in halide.

\section{Experimental Details}
Single crystals of Cs$_3$Cu$_2$Cl$_4$Br$_3$ were grown by slow evaporation from an equimolar aqueous solution of CsCl, CsBr, CuCl$_2$$\cdot$2H$_2$O and CuBr$_2$ at 40$^{\circ}$C. Thick square plate crystals with small mosaic spread were obtained. The typical dimensions of the single crystals are $1 \times 1 \times 0.5$ cm$^3$.

\begin{table}
\caption{Cs$_3$Cu$_2$Cl$_4$Br$_3$ crystal data.}
\label{table:1}
\begin{tabular}{@{\hspace{\tabcolsep}\extracolsep{\fill}}cc} \hline
Chemical formula & Cs$_3$Cu$_2$Cl$_4$Br$_3$  \\
Space group & $I4/mmm$  \\
$a$ ($\rm{\AA}$) & 5.2956(4)  \\
$c$ ($\rm{\AA}$) & 26.584(2)  \\
$V$ ($\rm{\AA}^3$) & 745.5(1)  \\
$Z$ & 2  \\
No. of observed reflections & 408 ($I$$>$3$\sigma$($I$)) \\
$R; wR$ & 0.075; 0.095  \\
\hline
\end{tabular}
\end{table}
The structural analysis of an as-grown crystal was performed at room temperature using a three-circle diffractometer equipped with a CCD area detector (Bruker SMART APEX). Monochromatic Mo-$K\alpha$ radiation was used as an X-ray source. Data integration and global-cell refinements were performed using data in the range of 2$\theta< 80^\circ$, and absorption correction was also performed. Structural parameters were refined by the full-matrix least-squares method using the TEXSAN program \cite{TEXSAN}. The crystal data are listed in Table I. The chemical formula was determined to be Cs$_3$Cu$_2$Cl$_4$Br$_3$ from the results of the structural analysis. Very weak reflections implying that the lattice symmetry should be primitive tetragonal were observed, resulting in a large $R$-factor ($=0.075$). Here, the space group of $I4/mmm$ for the average structure without weak satellite reflections was used for the structural analysis.

As shown in the next section, Cs$_3$Cu$_2$Cl$_4$Br$_3$ undergoes a structural phase transition at $T_{\rm s}\simeq330$ K, and the magnetic properties of the crystal once heated above $T_{\rm s}$ are different from those of the as-grown crystal. Therefore, to investigate the change in lattice symmetry, the X-ray powder diffraction patterns of the as-grown and heated crystals were obtained at room temperature with Cu-$K\alpha$ radiation.

Magnetic susceptibilities were measured using a SQUID magnetometer (Quantum Design MPMS XL). Specific heats were measured using a physical property measurement system (Quantum Design PPMS) by the relaxation method. ESR spectra were measured at room temperature using an X ($\sim 9$ GHz) band frequency.

\section{Results and Discussion}
\subsection{Crystal Structure}
\begin{figure}[htbp]
	\begin{center}
		\includegraphics[width=8cm,clip]{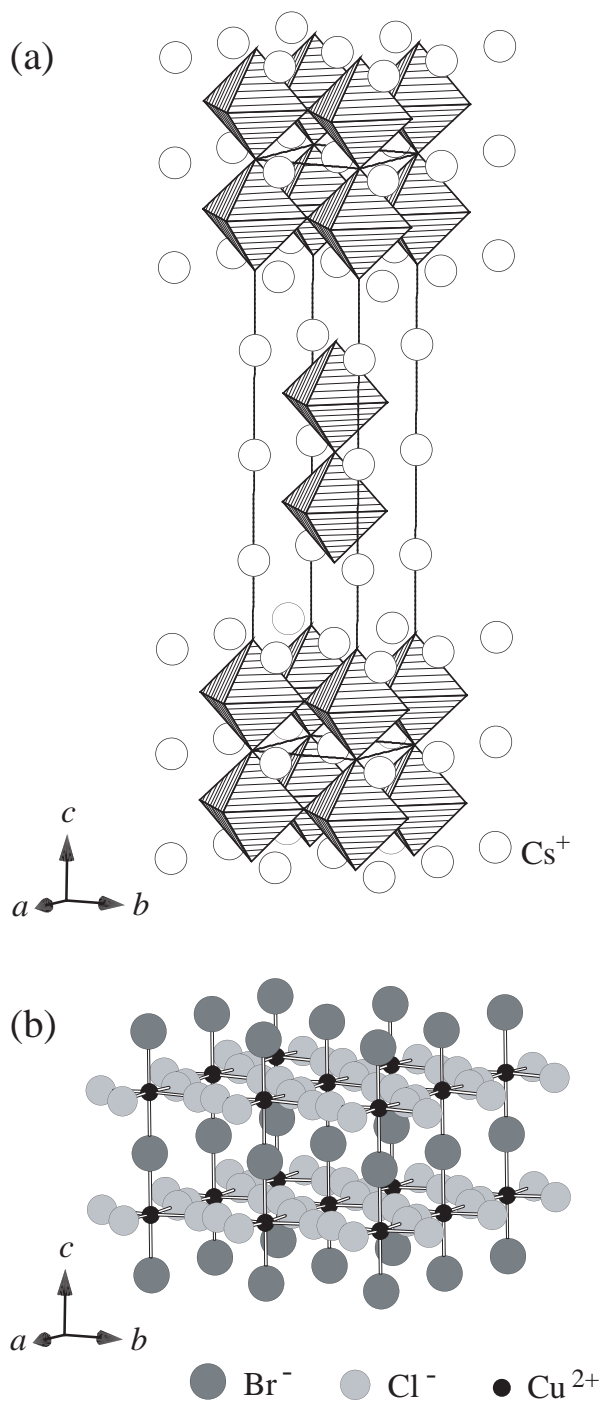}
	\end{center}
	\caption{Crystal structure of Cs$_3$Cu$_2$Cl$_4$Br$_3$. (a) Average structure given by $y=0.5$ for Cl, where hatched octahedra and open circles denote CuCl$_4$Br$_2$ octahedra and Cs$^+$ ions, respectively. (b) Double layer composed of CuCl$_4$Br$_2$ octahedra, in which two Cl sites with an occupancy of 0.5 are shown. Large and medium gray circles and small black circles denote Br$^-$, Cl$^-$ and Cu$^{2+}$ ions, respectively. Cs$^+$ ions are omitted in (b).
}
	\label{fig: fig1}
\end{figure}

The structure of an as-grown Cs$_3$Cu$_2$Cl$_4$Br$_3$ crystal is body-centered tetragonal, $I4/mmm$, with cell dimensions of $a=5.2956(4)$ \AA\ and $c=26.584(2)$ \AA, and $Z=2$.
Positional parameters are listed in Table II. The crystal structure of Cs$_3$Cu$_2$Cl$_4$Br$_3$ is shown in Fig. \ref{fig: fig1}. Disorder was observed for the positions of Cl$^-$ ions, which occupy the position 16($n$) with an occupancy of 0.5 on average. In Fig. \ref{fig: fig1}(a), we illustrate the average positions of Cl$^-$ ions given by $y=0.5$ for Cl. 

\begin{table}
\caption{Fractional atomic coordinates, equivalent isotropic displacement parameters ($\rm{\AA}^2$) and site occupancies.}
\label{table:2}
\begin{tabular}{lccccc}\hline
Atom  &  $x$  & $y$  & $z$ & $U_{\rm eq}$ & Occ.  \\ \hline
Cs(1) & 0 & 0 & 0.5 & 0.0240(3) & 1   \\
Cs(2) & 0 & 0 & 0.3224(7) & 0.0285(2) & 1  \\
Cu & 0 & 0 & 0.0943(4) & 0.0208(4) & 1  \\
Br(1) & 0 & 0 & 0  & 0.0285(5) & 1  \\
Br(2) & 0 & 0 & 0.1873(4) & 0.0366(4) & 1  \\
Cl  & 0 & 0.577(1) & 0.0916(2) & 0.021(1) & 0.5   \\ 
\hline
\end{tabular}
\end{table}
\begin{table}
\caption{Bond distances (\AA).}
\label{table:3}
\begin{tabular}{llccc}\hline
& Cu $-$ Br(1) & & 2.507(3) & \\ 
& Cu $-$ Br(2) & & 2.471(5) &  \\
& Cu $-$ Cl & & 2.240(6) &  \\
& Cu $-$ Cl$'$ & & 3.058(6) &  \\ 
\hline
\end{tabular}
\end{table}

The crystal structure of Cs$_3$Cu$_2$Cl$_4$Br$_3$ is closely related to the Sr$_3$Ti$_2$O$_7$ structure \cite{Sr3Ti2O7}. In the CuCl$_4$Br$_2$ octahedron, four Cl$^-$ ions and two Br$^-$ ions 
share corners along the $a\,(b)$-axis and the $c$-axis. CuCl$_4$Br$_2$ octahedra form double layers parallel to the $c$-plane, sharing their corners. All the double layers are equivalent and neighboring double layers are shifted by $({\mib a}+{\mib b})/2$. Figure \ref{fig: fig1}(b) illustrates a double layer composed of CuCl$_4$Br$_2$ octahedra, in which two Cl sites with an occupancy of 0.5 are shown. Interatomic distances are shown in Table III. Because of the Jahn-Teller effect and the unharmonic potential of the distortion modes and/or higher-order Jahn-Teller coupling \cite{Van, Opik, Liehr,Tanaka}, the octahedra are elongated along either of the two principal axes parallel to the $c$-plane. The octahedra are elongated antiferrodistortively in the $c$-plane, as observed in K$_2$CuF$_4$ \cite{K2CuF4}. 

Figure 2 illustrates the possible configurations of the hole orbitals $d(x^2-y^2)$ of neighboring Cu$^{2+}$ ions along the $c$-axis and the $a$- or $b$-axis in a double layer. The hole orbitals $d(x^2-y^2)$ are accompanied by the tetragonal elongation of CuCl$_4$Br$_2$ octahedra and arranged perpendicular to the elongated axes. The orbital configurations shown in (a)$-$(g) and (h)$-$(j) produce antiferromagnetic and ferromagnetic exchange interactions, respectively, through the $p$ orbitals of Br$^-$  and Cl$^-$ ions. Those shown in (a)$-$(e) lead to strong antiferromagnetic exchange interactions via straight exchange pathways.
In the as-grown crystal, the configuration of the hole orbitals of neighboring Cu$^{2+}$ ions along the $c$-axis in a double layer should be either of those shown in Figs. 2(a) and 2(b). Both orbital configurations can produce strong antiferromagnetic exchange interactions \cite{Kanamori}, as confirmed by the magnetic measurements shown below. The orbital configuration of neighboring Cu$^{2+}$ ions along the $a$- or $b$-axis in a double layer should be that shown in Fig. 2(i), because the elongation of the octahedra occurs antiferrodistortively in the double layer. The exchange interaction corresponding to this orbital configuration is weak and ferromagnetic \cite{Kanamori}, as observed in K$_2$CuF$_4$ \cite{Hirakawa}.

\begin{figure}[htbp]
	\begin{center}
		\includegraphics[width=7cm,clip]{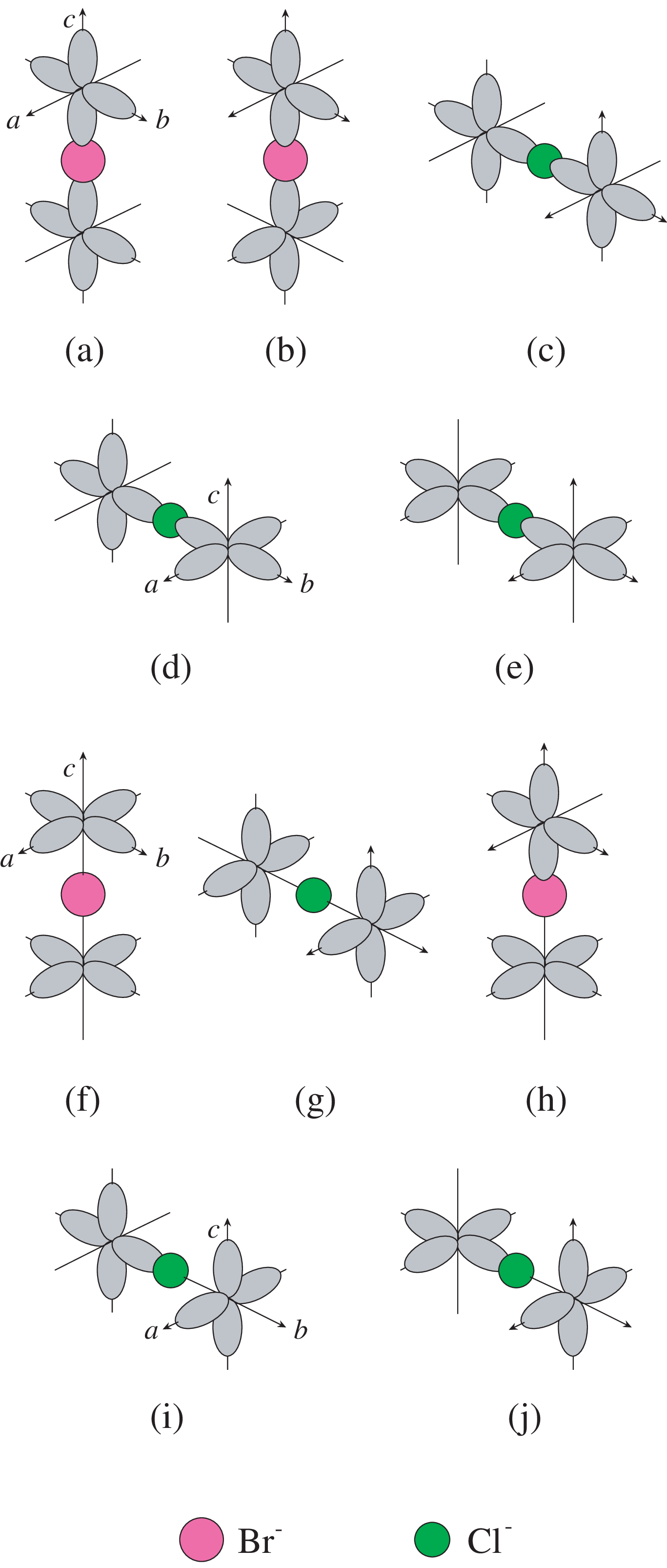}
	\end{center}
	\caption{Possible configurations of hole orbitals $d(x^2-y^2)$ of neighboring Cu$^{2+}$ ions along the $c$-axis and the $a$- or $b$-axis in a double layer. The elongation of CuCl$_4$Br$_2$ octahedra is assumed to occur owing to the Jahn-Teller effect and the unharmonic potential of the distortion modes and/or higher-order Jahn-Teller coupling. The hole orbitals $d(x^2-y^2)$ are arranged perpendicular to the elongated axes. The orbital configurations shown in (a)$-$(g) and (h)$-$(j) produce antiferromagnetic and ferromagnetic exchange interactions, respectively, through the $p$ orbitals of Br$^-$  and Cl$^-$ ions. Those shown in (a)$-$(e) lead to strong antiferromagnetic exchange interactions via straight exchange pathways.}
	\label{fig: fig2}
\end{figure}

\subsection{Magnetic Properties}
Figure 3 shows the temperature dependence of the magnetic susceptibility ${\chi}$ in Cs$_3$Cu$_2$Cl$_4$Br$_3$ measured for magnetic fields $H$ parallel and perpendicular to the $c$-axis. We used different samples for $H\parallel c$ and $H\perp c$. The measurement process is as follows: first, an as-grown crystal is cooled to 1.8 K. The susceptibility is measured upon heating to 390 K and then upon cooling to 1.8 K. The susceptibility of the as-grown crystal is negative below room temperature. 
With increasing temperature, the susceptibility displays a sudden increase at $T_{\rm s}\simeq330$ K, above which the susceptibility obeys the Curie-Weiss law. This anomaly is ascribed to a structural phase transition. Once a Cs$_3$Cu$_2$Cl$_4$Br$_3$ crystal is heated above $T_{\rm s}$, its magnetic susceptibility does not show anomalies at $T_{\rm s}$ with decreasing temperature. The susceptibility remains obeying the Curie-Weiss law with the Curie constant of $C=0.44$ emu$\cdot$K/mol and the Weiss constant of ${\Theta}=23$ K for $H\perp c$ and $C=0.47$ emu$\cdot$K/mol and ${\Theta}=1.3$ K for $H\parallel c$ down to 30 K. From the Curie constant, the number of Cu$^{2+}$ ions relevant to the susceptibility is estimated to be 50 \% for $H\perp c$ and 55 \% for $H\parallel c$. The number of relevant ions and the Weiss constant ${\Theta}$ are somewhat dependent on the sample. 
The structural phase transition at $T_{\rm s}$ is irreversible. However, the diamagnetic state is easily restored by the application of a hydrostatic pressure of $\sim$10 bar at room temperature. Thus, the paramagnetic state realized below  $T_{\rm s}\simeq330$ K after heating is the metastable state.

Figure 4 shows the enlargement of the negative region of the susceptibility observed for $T < T_{\rm s}$. The negative susceptibility is ascribed to the diamagnetism from core electrons. This result indicates that the magnetic ground state of the as-grown crystal is a spin singlet with an extremely large excitation gap. As mentioned in the previous subsection, the orbital configuration of neighboring Cu$^{2+}$ ions between layers is either of those shown in Figs. 2(a) and (b), both of which can produce strong antiferromagnetic exchange interactions. Thus, the singlet ground state of the as-grown crystal is attributed to the dimerization of spins between layers. 

Next, we evaluate the interlayer exchange interaction $J$ of the as-grown Cs$_3$Cu$_2$Cl$_4$Br$_3$ crystal using the following model:
\begin{eqnarray}
{\chi}=\frac{2g^2{\mu}_{\rm B}^2N}{k_{\rm B}T}\cdot \frac{1}{3+{\exp}\left(2J/k_{\rm B}T\right)} + \frac{C'}{T-{\Theta}} + {\chi}_0\ ,
\end{eqnarray}
where the first term is the susceptibility of noninteracting $N$ dimers, the second term is the Curie-Weiss term due to impurities or lattice defects and the last term is the temperature-independent susceptibility including the Van Vleck paramagnetism and the diamagnetism from core electrons. Here, the exchange interaction $J$ is defined as $2J({\boldsymbol S}_1\cdot {\boldsymbol S}_2$). Since no ESR signal is observed for the as-grown crystal, we use $g=2.05$ for $H\parallel c$ and $g=2.15$ for $H\perp c$, which are estimated from $g_{z}\simeq 2.25$ and $g_{x}\simeq 2.05$ obtained for $H$ parallel and perpendicular to the elongated axes of the CuCl$_6$ octahedron, respectively, in TlCuCl$_3$ and KCuCl$_3$ \cite{Oosawa1,Oosawa2}. The solid line in Fig. 4 denotes the fit obtained using eq. (1) with $J/k_B=995$ K, $C'=3.04\times 10^{-3}$ emu$\cdot$K/mol, ${\Theta}=-13.6$ K, and ${\chi}_0=-2.89\times 10^{-4}$ emu/mol for $H\parallel c$. For $H\perp c$, we obtain $J/k_B=984$ K, $C'=4.36\times 10^{-3}$ emu$\cdot$K/mol, ${\Theta}=-35.0$ K, and ${\chi}_0=-5.04\times 10^{-4}$ emu/mol. 
We observe that the orbital configuration shown in Fig. 2(a) or 2(b) produces an extremely strong antiferromagnetic exchange interaction of $J/k_B\simeq 1000$ K when mediated by Br$^-$ ions. The excitation gap is estimated to be ${\Delta}/k_{\rm B}=2J/k_{\rm B}\simeq 2000$ K. This large gap results in the diamagnetic state at room temperature for as-grown samples.
\begin{figure}
	\begin{center}
		\includegraphics[width=8cm,clip]{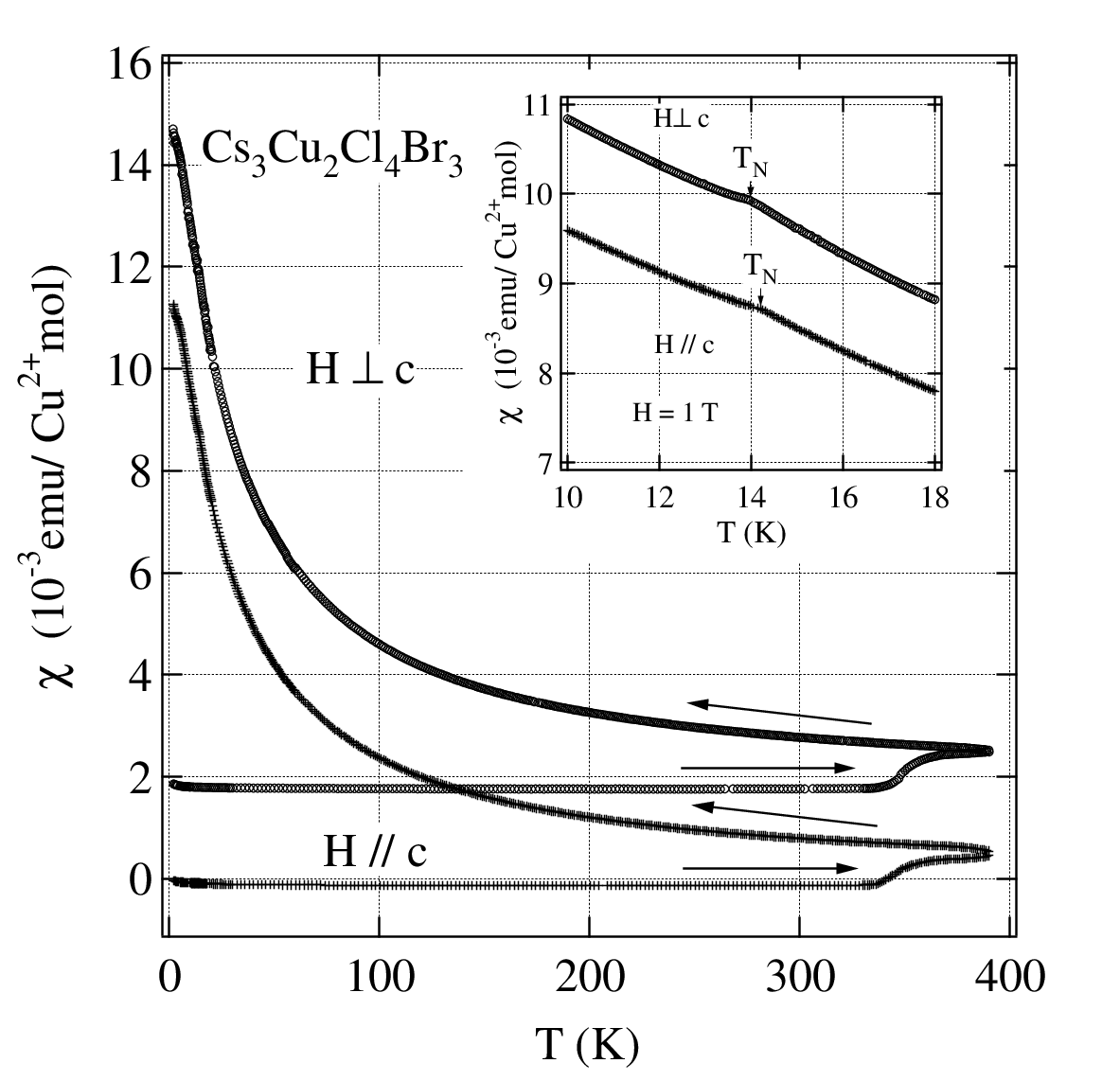}
	\end{center}
	\caption{Temperature dependence of magnetic susceptibility of Cs$_3$Cu$_2$Cl$_4$Br$_3$. The plot for magnetic field $H\perp c$ is shifted vertically by $2\times 10^{-3}$emu Cu$^{2+}$mol. Arrows denote the directions of temperature change. The inset shows the enlargement of susceptibility near the magnetic phase transition temperature of the heated crystal.}
	\label{fig: fig3}
\end{figure}
\begin{figure}
	\begin{center}
		\includegraphics[width=8cm,clip]{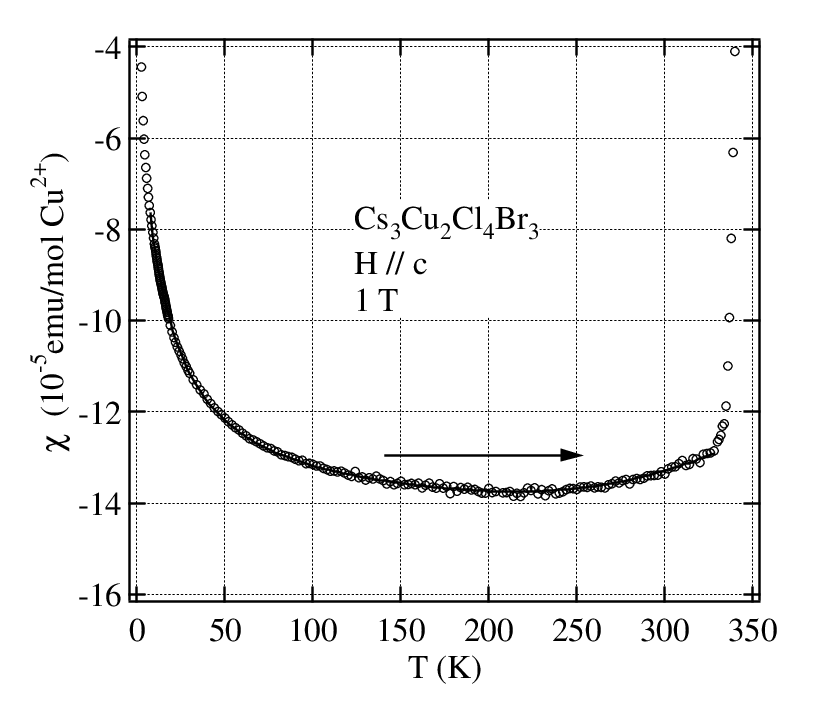}
	\end{center}
	\caption{Enlargement of temperature dependence of susceptibility of the as-grown Cs$_3$Cu$_2$Cl$_4$Br$_3$ crystal for $H\parallel c$. Arrow denotes the direction of temperature change. The solid line denotes the fit obtained using eq. (1) with parameters shown in the text.}
	\label{fig: fig4}
\end{figure}@

\begin{figure}
	\begin{center}
	\includegraphics[width=8cm,clip]{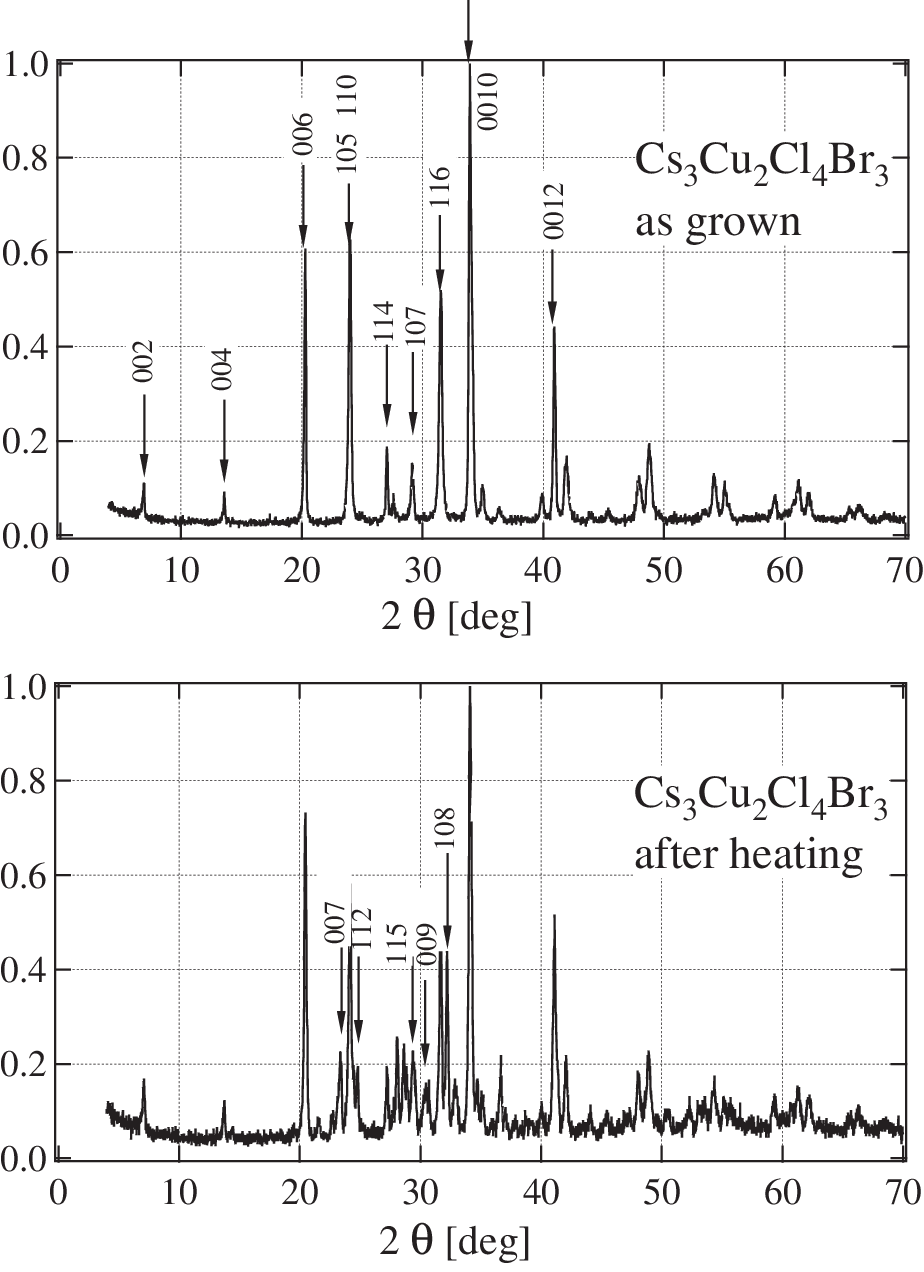}
	\end{center}
	\caption{X-ray powder diffraction patterns of (a) as-grown and (b) heated Cs$_3$Cu$_2$Cl$_4$Br$_3$ crystals.}
	\label{fig: fig5}
\end{figure}

To investigate the structural difference between the as-grown and heated crystals, we measure the X-ray powder diffraction patterns. The results are shown in Fig. \ref{fig: fig5}. For the as-grown crystal, only reflections indexed by $h+k+l=$ even number are observed, which is consistent with the space group $I4/mmm$. For the heated crystal, the lattice symmetry remains tetragonal. However, reflections indexed by $h+k+l=$ odd number are observed, in addition to those for $h+k+l=$ even number. This implies the lack of a body center. At present, the space group of the heated crystal is unclear.
@

\begin{figure}
	\begin{center}
		\includegraphics[width=8cm,clip]{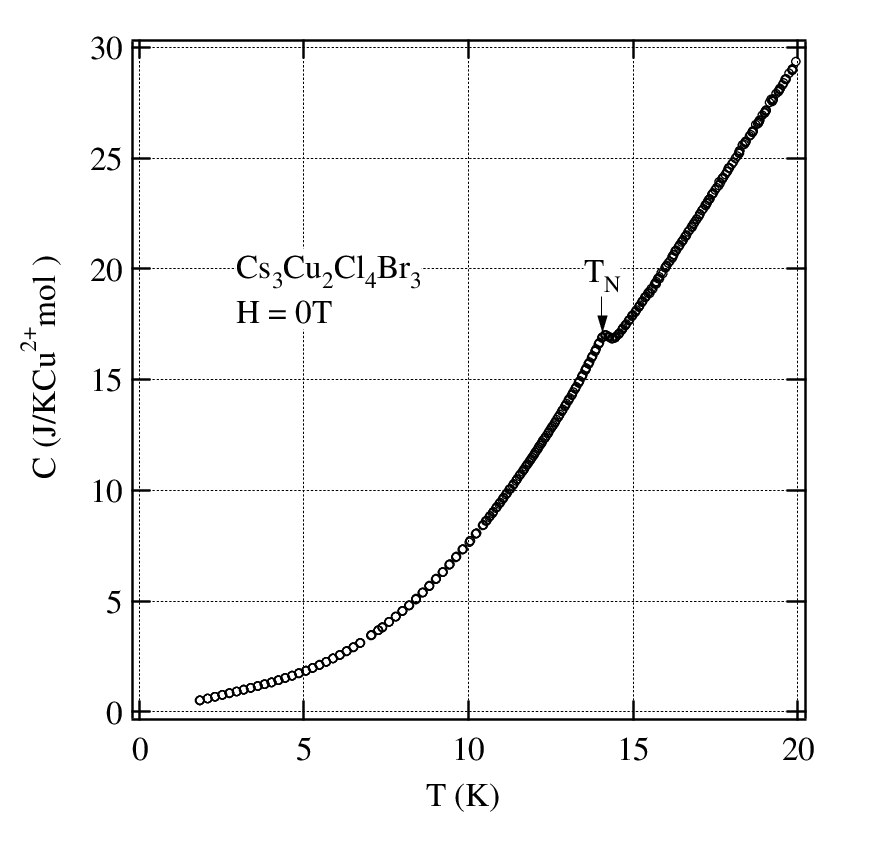}
	\end{center}
	\caption{Temperature dependence of total specific heat in Cs$_3$Cu$_2$Cl$_4$Br$_3$ after heating above $T_{\rm s}\simeq 330$ K. A magnetic phase transition is observed at $T_{\rm N}=14.0$ K.}
	\label{fig: fig6}
\end{figure}

As shown in the inset in Fig. 3, the susceptibility of the heated crystal exhibits a bend anomaly at $T_{\rm N}=14.0$ K, indicating magnetic ordering. Such magnetic ordering is also observed by specific heat measurement, as shown in Fig. \ref{fig: fig6}. The specific heat displays a small peak at $T_{\rm N}=14.0$ K, which is consistent with $T_{\rm N}$ determined from the susceptibility measurement. Since the exchange interaction in the as-grown crystal is $J/k_B\simeq 1000$ K, the magnetic specific heat should be small below room temperature. Then, we try to estimate the magnetic entropy $S_{\rm m}$ of the heated crystal, assuming the total specific heat of the as-grown crystal to be the lattice contribution of the heated crystal. However, we cannot successfully estimate $S_{\rm m}$. This is due to the lattice contributions of the as-grown and heated crystals having different temperature dependences at low temperatures. 

The results of susceptibility and specific heat measurements for the heated crystal indicate that the orbital configurations of the heated crystal differ from those of the as-grown crystal, in which hole orbitals lie only in the $ac$- and $bc$-planes, and their configurations along the $c$-axis and the $a$- or $b$-axis in a double layer are limited to those shown in Figs. 2(a), 2(b) and 2(i). For the heated crystal, an ESR signal with a linewidth of $1.2\sim 1.3$ kOe is observed. The $g$-factors obtained for $H\parallel c$ and $H\perp c$ are $g_{\parallel}=2.115(5)$ and $g_{\perp}=2.095(5)$, respectively. Their average $g_{\rm av}=(g_{\parallel}+2g_{\perp})/3\simeq 2.10$ is almost the same as $g_{\rm av}\simeq 2.12$ in TlCuCl$_3$ and KCuCl$_3$; however, it is smaller than $g_{\rm av}\simeq 2.22$ in A$_2$CuF$_4$ \cite{Sasaki}. Nevertheless, the difference between $g_{\parallel}$ and $g_{\perp}$ is small in the heated Cs$_3$Cu$_2$Cl$_4$Br$_3$ crystal. If the octahedra are elongated only along the $a$- and $b$-axes, as observed in the as-grown crystal, the $g$-factors should be $g_{\parallel}\simeq 2.05$ and $g_{\perp}\simeq 2.15$. Therefore, we can deduce that in the heated Cs$_3$Cu$_2$Cl$_4$Br$_3$ crystal, octahedra are elongated not only along the $a$- and $b$-axes, but also along the $c$-axis with almost the same probability. This implies that orbitals are in disorder.

As shown above, the paramagnetic susceptibility of the heated  Cs$_3$Cu$_2$Cl$_4$Br$_3$ crystal obeys the Curie-Weiss law. From the Curie constant, the number of spins relevant to the susceptibility is estimated to be $\simeq 50$\%. In the heated crystal, all the orbital configurations shown in Fig. 2 should be possible for neighboring Cu$^{2+}$ ions along the $c$-axis and the $a$- or $b$-axis in a double layer. The exchange interactions corresponding to the configurations in Figs. 2(a)$-$2(e) are antiferromagnetic and strong \cite{Kanamori}, and their magnitudes are expected to be $J/k_{\rm B}\sim 1000$ K. Hence, clusters composed of even number of spins that are coupled by such strong interactions cannot contribute to the paramagnetic susceptibility of the heated crystal. On the other hand, the exchange interactions for the configurations in Figs. 2(f)$-$2(j) are expected to be of the order of 10 K. Thus, clusters composed of spins coupled by such weak interactions can contribute to the paramagnetic susceptibility. From the results of the $g$-factors and magnetic susceptibility, we infer that the orbitals of Cu$^{2+}$ ions are in disorder once the crystal is heated above the structural phase transition temperature $T_{\rm s}\simeq330$ K and that the phase transition is an order-disoder transition of $d(x^2-y^2)$ orbitals.  

\section{Conclusions}
We determined the crystal structure of an as-grown Cs$_3$Cu$_2$Cl$_4$Br$_3$ crystal and investigated its magnetic properties. This compound has a double-layered perovskite structure of the Sr$_3$Ti$_2$O$_7$ type. In this as-grown crystal, octahedra are elongated along either of the two principal axes parallel to the $a$- and $b$-axes owing to the Jahn-Teller effect and the unharmonic potential of the distortion modes and/or higher-order Jahn-Teller coupling. A strong antiferromagnetic exchange interaction of $J/k_{\rm B}\simeq 1000$ K is produced between neighboring spins along the $c$-axis in the double layer through the orbital configuration shown in Figs. 2(a) or 2(b). This leads to a strong spin dimerization, resulting in the diamagnetic state at room temperature. With increasing temperature, Cs$_3$Cu$_2$Cl$_4$Br$_3$ undergoes a structural phase transition at $T_{\rm s}\simeq 330$ K. Once the crystal is heated above $T_{\rm s}$, its magnetic susceptibility obeys the Curie-Weiss law with a small Weiss constant. Magnetic ordering occurs at $T_{\rm N}=14.0$ K. In the heated crystal, octahedra are elongated not only along the $a$- and $b$-axes, but also along the $c$-axis with almost the same probability. This implies that orbitals are in disorder above $T_{\rm s}$ and that the structural phase transition at $T_{\rm s}$ is an order-disorder transition of $d(x^2-y^2)$ orbitals. 

\section*{Acknowledgment}
This work was supported by a Grant-in-Aid for Scientific Research and the 21st Century COE Program at Tokyo Tech ``Nanometer-Scale Quantum Physics'', both from the Ministry of Education, Culture, Sports, Science and Technology of Japan. 



\end{document}